\documentstyle[epsfig]{elsart}
\begin{document}
\vspace*{-1.3cm}\hspace*{\fill}
% \parbox{7.0cm}{ANL-PHY-8295-TH-96\\ {\sc to appear in Physics Letters B}}
\parbox{10.0cm}{\sc appears in Physics Letters B 379 (1996) 1--6}
\vspace*{+1.3cm}
\begin{frontmatter}
%....................... Title, Authors ...........................%
\title{Pomeron exchange and exclusive electroproduction 
of $\rho$-mesons in QCD} 
\author[anl,pitt]{M. A. Pichowsky,}
\author[anl]{T.-S.\ H.\ Lee}
\address[anl]{Physics Division, Argonne National Laboratory,
Argonne IL 60439-4843}
\address[pitt]{Department of Physics \& Astronomy, University of Pittsburgh,
Pittsburgh PA 15260}
%..................................................................%
\begin{abstract}
A Pomeron-exchange model of exclusive electroproduction of $\rho$-mesons is
examined using a dressed-quark propagator.  It is shown that by representing
the photon-$\rho$-meson-Pomeron coupling by a nonperturbative, confined-quark
loop, one obtains predictions for $\rho$-meson electroproduction that are in
good agreement with experiment.
% In this approach, a quark-Pomeron form factor is unnecessary.
\end{abstract}
\end{frontmatter}

%.....................................................................
\renewcommand{\-}{\!-\!}   %%removes some space around + & - signs.
\renewcommand{\+}{\!+\!}
\newcommand{\g}{\gamma}  
\newcommand{\grp}{$\gamma \rho P$ }
\newcommand{\sfrac}[2]{\mbox{\footnotesize $\frac{#1}{#2}$}}
\sloppy
%.....................................................................
{\em 1.~Introduction.}  It is observed that high energy hadron elastic cross
sections are forward peaked and rise slowly with energy $\sqrt{s}$.  This is
attributed to the exchange of a phenomenological object known as the Pomeron.
Attempts to describe
the Pomeron as the exchange of two nonperturbatively-dressed gluons between
quarks typically treat the quarks as free (i.e.~all quark lines are
represented by free, constituent-quark propagators).  However, it has been
argued that quark confinement is necessary to obtain a reasonable description
of the Pomeron in terms of a gluon exchange \cite{ButtnerPennington}.  In
this work we investigate the role of confined-quark loops in a
Pomeron-exchange model.

We examine a model, a modification of that in Ref.~\cite{DL84}, in which the
Pomeron mediates the long range interaction between a confined quark and a
nucleon.  Our model provides a framework in which some nonperturbative
aspects of Pomeron exchange can be investigated.  We apply the model to, and
present predictions for, diffractive $\rho$-meson electroproduction cross
sections and compare them to recent data from the ZEUS Collaboration
\cite{ZeusPhoto} and previous measurements \cite{CHIOEMC,Data} in a wide
range of energy and photon momentum squared.

Herein we illustrate that, in $\rho$-meson electroproduction, agreement with
experiment can be obtained by including the full effect of the quark loop,
using a finite-width $\rho$-meson Bethe-Salpeter amplitude.  We find that,
even at high energies, the inclusion of a confined-quark loop has
considerable impact on such exclusive processes.

{\em 2.~A model for Pomeron exchange.}  In the model of Ref.~\cite{DL84}, the
Pomeron-nucleon coupling is described by the vertex $ F_{\mu}(t) \equiv 3
\beta_0 \g_{\mu} f(t)$, where $-t$ is the Pomeron momentum squared, 
$f(t)$ is a phenomenologically determined form factor, $\beta_0$ is a
coupling strength, 
and the factor of 3 counts the number of quarks in the nucleon.  With this
vertex, the $pp$ elastic scattering amplitude due to single Pomeron exchange
is
\begin{eqnarray}
T_{pp\rightarrow pp} &=& i \bar{u}(p_3) F_{\mu}(t) u(p_1) 
\,  G_{\cal P}(s,t)  
%%\nonumber \\ & & \times 
\; \bar{u}(p_4) F_{\mu}(t) u(p_2),
\label{Tpp}
\end{eqnarray}
where $G_{\cal P}(s,t)$ is assumed to have the form 
\begin{equation}
G_{\cal P}(s,t) = (- i \alpha^{\prime} s)^{\alpha(t) - 1} 
\>,  \label{PomProp}
\end{equation}
with $\alpha(t) = 1 + \epsilon + \alpha^{\prime} t$.  
The parameters $\epsilon$,
$\alpha^{\prime}$, and $\beta_0$, are fixed by requiring that
$T_{pp\rightarrow pp}$ 
reproduces the observed differential and total $pp$ elastic scattering cross
sections.  It was shown in Ref.~\cite{DL84} that with $f(t) = F_1(t)$,
the isoscalar nucleon electromagnetic form factor, this model is able to
describe the large body of $pp$ and $\bar{p}p$ elastic scattering data.  In
this model the Pomeron couples to hadrons as an isoscalar photon.  
Throughout this work,
following Ref.~\cite{DL87}, we use $\epsilon=0.08$,
$\alpha^\prime=0.25$~GeV$^{-2}$, $\beta_0=2.0$~GeV$^2$ and $F_1(t)\equiv (4
M_N^2 - 2.8 t) / (4 M_N^2 - t) / (1 - t /.7)^2$.

In employing this model in a calculation of the small-$x$ behavior of the
proton structure function a form factor ($\mu_0^2 / (\mu_0^2 + p^2)$, $\mu_0
=$ 1.2~GeV) was introduced, for each off-shell quark leg of momentum $p$ that
couples to the Pomeron, in order to ensure convergence of the quark triangle
diagram used to describe the $\g^\ast\g^\ast P$ vertex~\cite{DL87}.  In the
application of this model to exclusive $\rho$-meson electroproduction, the
same form factor is employed in the quark triangle diagram that describes the
photon-$\rho$-meson-Pomeron ($\g\rho P$) coupling.  Using an ``on-shell
approximation'' and a free, constituent-quark propagator with quark mass $M_q
= \sfrac{1}{2} m_{\rho}$, a value of the loop integral is inferred from the
$\rho \rightarrow e^+e^-$ electromagnetic decay constant.  In our calculation,
good agreement with the electroproduction data is obtained without such a
form factor.

\begin{figure}[t]
\centering{\
   \epsfig{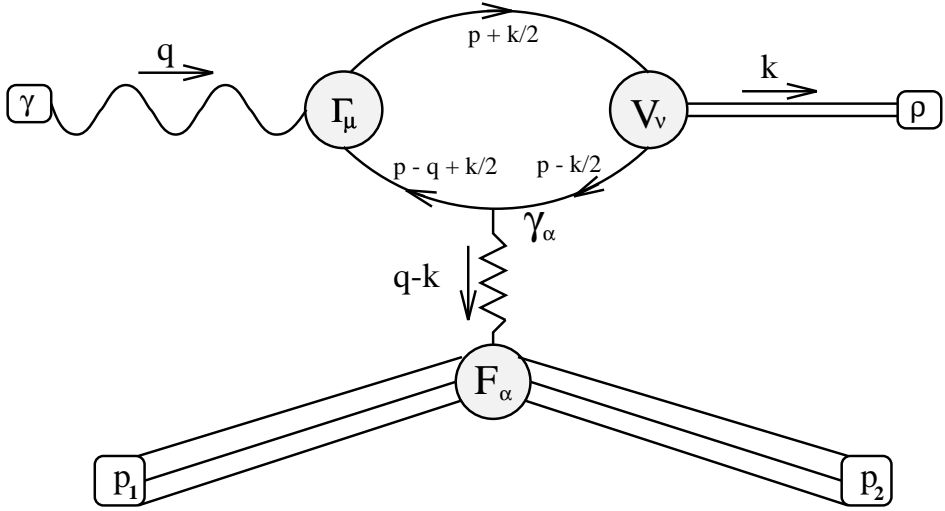} }
\caption{Lowest order diagram for the $\rho$-meson electroproduction current 
given by Eq.~(\protect\ref{Current}).} 
\label{Fig:Loop}
\end{figure}

{\em 3.~A model for \grp coupling.}  We employ a field theoretic framework,
formulated in Euclidean space\footnote { We use the metric $\delta_{\mu \nu}
= {\rm diag}(1,1,1,1)$ and Hermitian, traceless, Dirac matrices which satisfy
$\{\g_{\mu},\g_{\nu}\} = 2 \delta_{\mu \nu}$.  },
in which the impulse approximation to the $\rho$-meson electroproduction
current matrix element is
\begin{equation}
\langle p_2 m_2;k \lambda_{\rho} | J_{\mu} | p_1 m_1 \rangle = 
2 \beta_0 t_{\mu \alpha \nu}(q,k) \,
\varepsilon_{\nu}^{\lambda_{\rho}}(k)  \, G_{\cal P}(\bar{w}^2,t) 
\, \bar{u}(p_2) F_{\alpha}(t) u(p_1) ,
\label{Current}
\end{equation}
which is illustrated in Fig.~\ref{Fig:Loop}.  
Here $u(p_1)$ and
$\bar{u}(p_2)$ are incoming and outgoing nucleon spinors, 
$\varepsilon_{\nu}^{\lambda_{\rho}}(k)$ is the polarization vector of the
$\rho$-meson, and $t = - (p_1 - p_2)^2 \leq 0$.  
The explicit factor of 2 arises because of the equivalence
under charge conjugation, $C$, of the two leading order contributions, in
which the Pomeron strikes the quark or antiquark.  In this way we are
naturally led to a ``quark counting rule'' for the \grp vertex function,
$t_{\mu \alpha \nu}(q,k)$.  The \grp vertex in Eq.~(\ref{Current}) is
analogous to the nucleon-Pomeron vertex, $F_{\alpha}(t)$, from $pp$
scattering, in the sense that it represents the Pomeron-hadron coupling, and
in our model is given by
\begin{equation}
t_{\mu \alpha \nu}(q,k) = \frac{3 e_0}{(2\pi)^4} {\rm tr} \! \int \! \! 
d^{4}\!p S(p_+) \Gamma_{\mu}(p_+,p_+\-q) 
 S(p_+\-q)   \g_{\alpha}  S(p_-)   
V_{\nu}(p_-,p_+),
\label{tman}
\end{equation}
where $p_{\pm} = p \pm \sfrac{1}{2} k$.
In Eq.~(\ref{tman}) $S(p)=[i\gamma\cdot p A(p^2) + B(p^2)]^{-1}$ is the
dressed-quark propagator, $\Gamma_{\mu}(p,p^{\prime})$ is the dressed
photon-quark vertex, and $V_{\nu}(p,p^{\prime})$ is the $\rho$-meson
Bethe-Salpeter amplitude, which will be discussed below. The trace is over
Dirac indices.

%%%.......................................................Figure 1:
%%%.......................................................Figure 1:

In Fig.~\ref{Fig:Loop} the square of the quark-nucleon energy is 
$w^2 = -(p_1 + q - \sfrac{1}{2} k - p)^2$, which depends on the loop
integration variable $p$.  
To make contact with Refs.~\cite{DL87} we use an average value
$\bar{w}^2 = -(p_1 + q - \sfrac{1}{2} k )^2$ which allows
one to write the electroproduction amplitude as a product of Pomeron-hadron
vertices and Pomeron-exchange amplitude, $G_{\cal P}(\bar{w}^2,t)$.
We employ this ``factorization Ansatz'' throughout this work.
In our framework, the implication of this Ansatz can be explored and 
will be discussed elsewhere.

As remarked above, herein each of the terms appearing in the hadron current,
$J_{\mu}$ in Eq.~(\ref{Current}), is taken from Refs.~\cite{DL87} except the
\grp vertex, $t_{\mu \alpha \nu}(q,k)$, a calculation of which is the focus
of this work.

\begin{figure}[t]
\centering{\
 \epsfig{figure=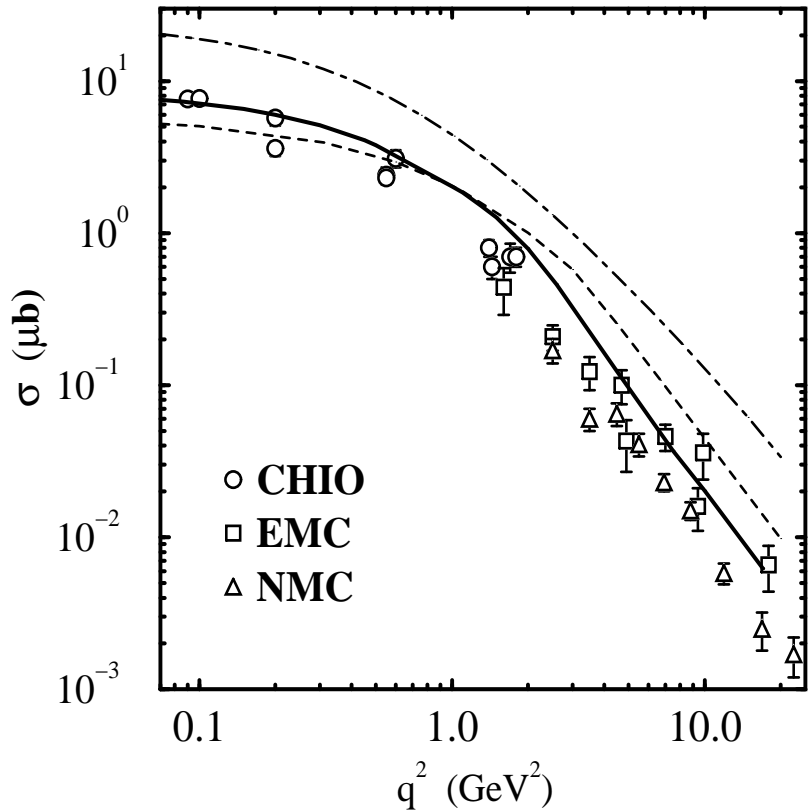,height=11.3cm,rheight=11cm,width=13.0cm}  }
\caption{
The total $\rho$-meson electroproduction cross section.
The solid line is our model, the dash-dotted line is obtained using the 
``on-shell approximation'',  
and the dashed line is our model with a simplified dressed-quark propagator.
The data are from Refs.~\protect\cite{CHIOEMC}
and error bars shown are statistical.
The curves are calculated at $W = 10$ GeV, and $\epsilon = 0.85$ 
(average value given by EMC in Ref.~\protect\cite{CHIOEMC}).}
\label{Fig:Qsq}
\end{figure}

The ``on-shell approximation'' of Refs.~\cite{DL87} can be reproduced in our
approach by taking the $\rho$-meson BS amplitude to be
\begin{equation}
V_{\nu}(p\-\sfrac{1}{2}k,p\+\sfrac{1}{2}k) 
= (2\pi)^4 \delta^4\!(p) \frac{m_{\rho}^4}{N_{\rho}} 
\g_{\nu}
\label{OnShellRhoV}
\>.
\end{equation}
Using a bare photon vertex $\Gamma_{\mu} = \g_{\mu}$, free constituent-quark
propagator $S(p) = (i \g \cdot p + \sfrac{1}{2} m_{\rho})^{-1}$, and the
above form of $\rho$-meson amplitude, the loop integration in Eq.~(\ref{tman})
can be performed trivially.  The only unknown parameter is the $\rho$-meson
vertex normalization, $N_{\rho}$, which is fixed by calculating the $\rho
\rightarrow e^+ e^-$ decay width, in the same approximation.  In
Fig.~\ref{Fig:Qsq} we compare the calculated results obtained in this way
with the available data.

In our calculation of the loop integral in Eq.~(\ref{tman})
we use the  photon-quark vertex, $\Gamma_{\mu}$, and quark propagator,
$S$, developed in studies of QCD based on the Dyson-Schwinger equations
(DSEs) in Euclidean space.  
(A review can be found in Ref.~\cite{RobertsWilliams94}.)
DSE studies have been employed successfully in the study of hadron and
electromagnetic interactions of pseudoscalar and vector mesons.
In the study of vector mesons the simple Gaussian Ansatz for the $\rho$-meson
BS amplitude 
\begin{equation}
V_{\nu}(p\-\sfrac{1}{2}k,p\+\sfrac{1}{2}k) =
\left( \g_{\nu} + \frac{k_{\nu} \g \cdot k}{m^2_{\rho}} \right)
\frac{e^{-p^2/\Lambda^2} }{N_{\rho}},
\label{RhoV}
\end{equation}
has proven phenomenologically successful \cite{Tandy}.  In Eq.~(\ref{RhoV}),
$k$ is the center-of-mass momentum of the $\rho$-meson and $p$ is the
quark-antiquark relative momentum.  The normalization, $N_{\rho}$, is fixed
by the canonical Bethe-Salpeter normalization and the range, $\Lambda$, is
determined by requiring that the model reproduce the $\rho \rightarrow e^+
e^-$ decay width.

The essential elements of our calculation of both the \grp vertex and $\rho
\rightarrow e^+ e^-$ width are the quark propagator and photon-quark vertex.

The photon-quark vertex, $\Gamma_{\mu}$, has been studied in some detail within
the DSE approach \cite{RobertsWilliams94}.
The quark propagator and photon-quark vertex must be dressed in a
consistent manner in order to satisfy the Ward-Takahashi identity (WTI).
In this work we use the vertex \cite{BallChiu}, 
\begin{equation}
i \Gamma_{\mu}^{\rm BC}(p,p^{\prime}) = 
\sfrac{i}{2} \g_{\mu} f_1(p^2,p^{\prime 2}) 
+ (p+p^{\prime})_{\mu} [i \g \cdot \sfrac{p+p^{\prime}}{2}
f_2(p^2,p^{\prime 2}) + f_3(p^2,p^{\prime 2})],
\label{GammaBC}
\end{equation}
with $f_1(p^2,p^{\prime 2})= A(p^2) + A(p^{\prime 2})$, 
$f_2(p^2,p^{\prime 2}) = (A(p^2) - A(p^{\prime 2}))/(p^2 - p^{\prime 2})$,
and  $f_3(p^2,p^{\prime 2}) = (B(p^2) - B(p^{\prime 2}))/(p^2 -
p^{\prime 2})$.  
This vertex satisfies the WTI
and transforms correctly under $C$, $P$, $T$, and Lorentz transformations.
It has the correct perturbative limit and is free of kinematic singularities.
However, it is not unique. 
The addition of transverse terms is important in connection with true
gauge covariance and multiplicative renormalizability, 
but contributes little or not at all to physical observables \cite{Craig}. 

The dressed-quark propagator,
$S(p) = - i \g \cdot p \sigma_V(p^2) + \sigma_S(p^2)$:
\begin{eqnarray}
\bar{\sigma}_S(x) &=& \frac{1\-e^{-b_1x}}{b_1 x} \frac{1\-e^{-b_3 x}}{b_3 x} 
(b_0 + b_2 \frac{1\-e^{-b_4 x}}{b_4 x} ) 
+ \bar{m}\frac{1 \- e^{-2(x+\bar{m}^2)}}{x + \bar{m}^2 },
\label{QuarkProp}\\
\bar{\sigma}_V(x) &=& 
\frac{2(x+\bar{m}^2) - 1 + e^{-2(x+\bar{m}^2)} }{ 2(x+\bar{m}^2)^2 },
\nonumber
\end{eqnarray}
with $x = p^2/\lambda^2$, $\bar{\sigma}_{S} = \lambda \sigma_S$,
$\bar{\sigma}_V = \lambda^2 \sigma_V$.
The qualitative features of this algebraic form follow from 
numerous studies of quark-DSE using realistic model forms for the gluon
propagator and quark-gluon vertex.  The parameters are fixed in
Ref.~\cite{Roberts94} by requiring that the model provide a good description
of the pion; e.g.\ $f_{\pi}$, $\pi$-$\pi$ scattering lengths, pion charge
radius, and form factor:
$b_0 = 0.118$, $b_1 = 2.51$, 
$b_2 = 0.525$, $b_3 = 0.169$, and $b_4 = 10^{-4}$.
In Eq.~(\ref{QuarkProp})
$m = \lambda \bar{m} = 6.7$ MeV  
is the bare quark mass and $\lambda =
0.568 $ GeV is the momentum scale. 
This dressed-quark propagator has no Lehmann representation and hence can be
interpreted as describing a confined particle since this feature is
sufficient to ensure the absence of quark production thresholds in 
$t_{\mu \alpha \nu}(q,k)$.
With this choice of quark propagator,
the Ball-Chiu vertex in Eq.~(\ref{GammaBC}) and $\rho$-meson Bethe-Salpeter
amplitude in Eq.~(\ref{RhoV}) 
are fixed and there are {\em no} free parameters.  Using this dressed-quark
propagator the $\rho \rightarrow e^+ e^-$ decay width is reproduced with
$\Lambda = 0.495 $ GeV.

%.....................................................................
{\em 4.~Results and discussion.}  We calculate the exclusive $\rho$-meson
electroproduction cross section, which can be separated into a transverse
term and a longitudinal term: $d\sigma / d\Omega_e^\prime dE_e^\prime =
\Gamma (\sigma_T(q^2,W) + \epsilon \sigma_L(q^2,W) )$ where $\Gamma$ and
$\epsilon$ are the usual virtual photon flux and photon polarization
parameter, respectively \cite{Akerlof}.  The transverse and longitudinal
cross sections depend only on the invariant mass, $W^2 =-(p_2 + k)^2$, and
the space-like momentum squared of the virtual photon, $q^2$.  We define the
Hand photon momentum $K_H = (W^2 - M_N^2)/(2 M_N)$, and obtain (in the
electron scattering plane $\phi = 0$),
\begin{eqnarray}
\frac{d\sigma_T}{d\Omega} &=&  \frac{1}{(2\pi)^2} \frac{M_N}{4 W}
\frac{|\vec{k}|}{K_H}  \frac{1}{2} \sum_{\rm spins}
(J_xJ_x^{\dag}+J_y J_y^{\dag})_{\phi = 0} 
, \nonumber \\
\frac{d\sigma_L}{d\Omega} &=& \frac{1}{(2\pi)^2} \frac{M_N}{4 W}
\frac{|\vec{k}|}{K_H}  \frac{q^2}{\omega^2}
\sum_{\rm spins}  (J_zJ_z^{\dag} )_{\phi = 0} 
\label{CrossSections}
\>.
\end{eqnarray}
Here $J_{\mu}$ is the electromagnetic hadron current defined in
Eq.~(\ref{Current}),
$M_N$ is the nucleon mass and $\omega$ is the photon momentum in the
center-of-momentum frame.

%%.......................................................Figure 2:
%%.......................................................Figure 2:

In Fig.~\ref{Fig:Qsq} we compare our predictions of the total cross sections
for exclusive $\rho$-meson electroproduction with the data from the CHIO, EMC
and NMC Collaborations \cite{CHIOEMC}, in the energy region 5.5 $ < W < $ 16
GeV.  The result of our model using the $\rho$-meson amplitude,
Eq.~(\ref{RhoV}), the Ball-Chiu vertex, Eq.~(\ref{GammaBC}), and  
confined-quark propagator, Eq.~(\ref{QuarkProp}) is shown as a solid curve
in Fig.~\ref{Fig:Qsq}.  It is in good agreement with the data.  No parameters
were varied to achieve this result.  The small-$q^2$ behavior is tightly
constrained and determined by the fact that the quark-photon vertex satisfies
the Ward identity~\cite{Roberts94}.  The dash-dotted curve is obtained by
employing the ``on-shell approximation'' (using the $\delta$-function form of
$\rho$-meson amplitude, Eq.~(\ref{OnShellRhoV}), and a free, constituent-quark
propagator).  This is our estimation of the result that would be obtained in
the model of Refs.~\cite{DL87} if no quark-Pomeron vertex is introduced.  It
overestimates the data.  We can fit the data if we introduce a quark-Pomeron
form factor $\mu_0^2 / (\mu_0^2 + p^2)$, with $\mu_0 = 1.2$ GeV.  Our
calculation suggests that this form factor models the quark substructure of
the \grp vertex amplitude and that the cross section in exclusive processes
is sensitive to the Bethe-Salpeter amplitude of the final-state meson.
Therefore an analysis of the data using the ``on-shell approximation'' in the
model of Refs.~\cite{DL87} may be misleading.

%%.....................................................Figure 3:
\begin{figure}[t] 
\centering{\
 \epsfig{figure=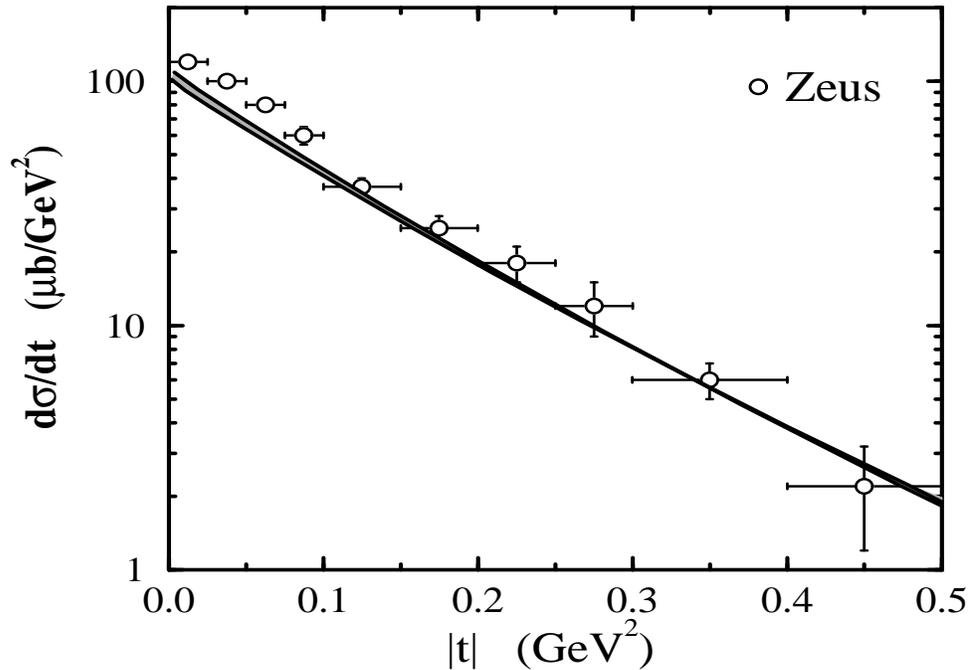,height=10.0cm,rheight=9.0cm,width=13.8cm}  }
\caption{
The differential cross section for exclusive $\rho$-meson photoproduction 
($q^2 = 0$) obtained by the ZEUS Collaboration 
over a range of energies $60 < W < 80$ GeV \protect\cite{ZeusPhoto}.
The region between the two solid curves is the prediction of this work 
for the same range of energies.
}
\label{Fig:dsigt}
\end{figure}
%%.....................................................Figure 3:

Our prediction for the differential cross section of photoproduction 
of $\rho$-mesons at $60 < W < 80$ GeV (and $q^2 = 0$) 
is also in good agreement with the recent data from ZEUS
\cite{ZeusPhoto}, as shown in Fig.~\ref{Fig:dsigt}.

\begin{figure}[t]
\centering{\
\epsfig{figure=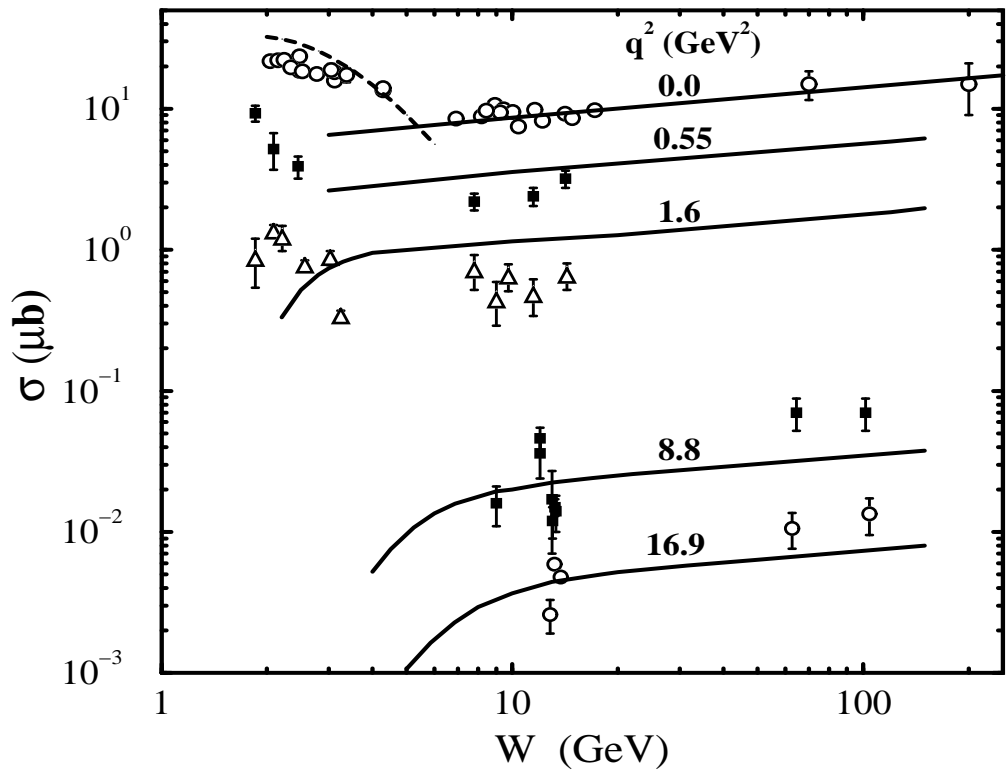,height=11cm,rheight=10cm,width=14.4cm}  }
\caption{
Energy dependence of the total $\rho$-meson electroproduction cross section
at several values of $q^2$. 
The solid line is this work and the dashed line is a meson exchange
calculation~\protect\cite{NozawaLee}. 
The data shown are from Refs.~\protect\cite{ZeusPhoto,CHIOEMC,Data}.
Error bars shown are statistical.
}
\label{Fig:sigmaW}
\end{figure}

The energy range in Fig.~\ref{Fig:dsigt} is $60 < W < 80 $ GeV, while $W =
10$ GeV in Fig.~\ref{Fig:Qsq}.  This emphasizes that the Pomeron
phenomenology is successful over a large energy domain.  This point is
illustrated again in Fig.~\ref{Fig:sigmaW}, in which, as expected from the
form of $G_{\cal P}(s,t)$, the cross sections rise very slowly with $W$.  The
rate of this increase is determined by the value of $\epsilon$ in
Eq.~(\ref{PomProp}).

%%.......................................................Figure 4:
%%.......................................................Figure 4:

In the topmost curve in Fig.~\ref{Fig:sigmaW} ($q^2 =  0$) 
Pomeron exchange dominates for $W > 6$ GeV.  
For $W < 6$ GeV the Pomeron contribution alone does not provide for
agreement with the data.  This is consistent with the expectation that, on
this small-$W$ domain, other mechanisms, such as quark or meson exchanges,
provide a significant contribution.
As an illustration of this point we include the results (dashed curve in
Fig.~\ref{Fig:sigmaW}) for $\rho$-meson photoproduction ($q^2= 0$)
calculated using the meson-exchange model of Ref.~\cite{NozawaLee}.

As discussed above, the photon-quark vertex and $\rho$-meson amplitude are
completely determined once the dressed-quark propagator is specified.
To explore the model sensitivity we have also calculated the $\rho$-meson
electroproduction cross section using the simplified dressed-quark
propagator: 
$S(p) = (1 - \exp[-(1 + p^2/M^2)])/(i\gamma\cdot p + M)$,
where $M$ is an effective-mass parameter.  This model preserves the
realization of quark confinement described above but is inadequate for a
description of low energy pion properties.  With $M = 0.5$ GeV, $\Lambda =
0.465$ GeV reproduces the $\rho \rightarrow e^+e^-$ width and the
electroproduction cross section determined in this case is the dashed curve
in Fig.~\ref{Fig:Qsq}.  The shape and magnitude of this curve are insensitive
to $M$ on the range $ 0.4 < M < 0.7$ GeV.  Comparing the solid and dashed
curves in Fig.~\ref{Fig:Qsq}, we find that our conclusion concerning the
importance of the loop integration does not depend on the exact form of quark
propagator employed.  Quantitative agreement with data requires the form of
Eq.~(\ref{QuarkProp}).

%.....................................................................

{\em 5.~Conclusion. }  We have calculated $\rho$-meson electroproduction
cross sections with the photon-$\rho$-Pomeron coupling described by a
confined-quark loop with a finite-width $\rho$-meson Bethe-Salpeter
amplitude.  Good agreement with the available data is obtained without
employing a form factor at the quark-Pomeron vertex.  This does not entail,
however, that the Pomeron is pointlike because information about its
non-pointlike nature may be contained in the parameterization of the
Pomeron-exchange amplitude, $G_{\cal P}(s,t)$.

%.....................................................................
\medskip

The authors are indebted to C.~D.~Roberts for many fruitful discussions.
This work is supported by the U.~S. Department of
Energy, Nuclear Physics Division, under contract W-31-109-ENG-38.
M.~A.~P. is also supported by the Division of Educational Programs of Argonne
National Laboratory and the Dean of Graduate Studies of the Faculty of Arts
and Sciences of the University of Pittsburgh. 

%.....................................................................

\end{document}